\documentclass[12pt]{article}
\usepackage{titling}
\usepackage[utf8]{inputenc}
\usepackage{amsmath}
\usepackage{amsthm}
\usepackage{mathtools}
\usepackage{amssymb}
\usepackage{mathrsfs}
\usepackage[english]{babel}
\usepackage{changepage}
\usepackage{graphicx}
\usepackage{mathrsfs}
\usepackage{relsize}
\usepackage{url}
\usepackage{breakurl}
\usepackage[breaklinks]{hyperref}
\usepackage{cite}
\usepackage{authblk}
\usepackage{appendix}
\usepackage{accents} 
\graphicspath{ {./images/} }
\usepackage{booktabs,makecell}
\usepackage{subcaption}

\usepackage{fullpage}

\usepackage[normalem]{ulem}
\usepackage[dvipsnames]{xcolor}

\begin{document}

\title{Inaccessibility in public transit networks: A study of
the London and New York City subway networks}
\author{\normalsize Katherine Betz$^1$}
\date{%
    \footnotesize$^1$ Department of Mathematics, State University of New York at Buffalo, NY 14260-2900, USA\\[1ex]}
\maketitle

\begin{abstract}
The study of networks derived from infrastructure systems has received considerable attention, yet the accessibility of such systems, particularly public transit networks, remains comparatively underexplored. This study examines infrastructure-based accessibility in two major legacy transit systems, the London Underground and the New York City Subway. We construct accessibility-restricted subnetworks in which nodes represent accessible stations and edges represent connections between them, and compare them with the corresponding full transit networks. We analyze degree distributions, the cyclomatic number, betweenness centrality, and closeness centrality to identify key accessible hubs and differences in network organization. London has a larger share of accessible stations, and its accessible network retains more of the alternative routes of its full network, whereas New York City has a more compact accessible core concentrated around central Manhattan and major interchanges. Comparing the observed accessible stations with random sets of the same size, we find that in London they have higher betweenness centrality than random allocations, but more of their stations are beyond two stops of an accessible station, consistent with accessibility concentrated at major interchanges and along fully step-free lines. In New York City, their higher betweenness centrality is fully explained by their concentration at interchanges, and their coverage is similar to that of random allocations. We also examine income and daytime activity around accessible stations. In both systems, several of the most central accessible stations are in high-activity locations, whereas accessible stations show no consistent relationship with median income. Our findings highlight substantial differences in accessibility between and within the two systems and demonstrate the utility of network-theoretic methods for understanding and improving transit infrastructure.
\end{abstract}

\textbf{keywords}: infrastructure, accessibility, network science

\section{Introduction}
Public transit systems are central components of urban infrastructure worldwide, serving daily metropolitan movement as well as long-distance travel.  As the priorities of transportation planning shift from speed and reduced travel times towards system reliability and low environmental impact, accessibility is becoming a crucial metric for informing policy, planning, and decision-making.  While many transit systems function effectively for able-bodied riders, individuals with disabilities frequently encounter substantial barriers within the transit infrastructure. According to a 2022 survey conducted by the Centers for Disease Control and Prevention, approximately 1 in 4 Americans, which is over 70 million people, report having a disability \cite{CDC24}.  In a separate 2022 survey, the Bureau of Transportation Statistics estimated that 18.6 million Americans have a travel-limiting disability, defined as “a temporary or permanent condition or handicap that makes it difficult to travel outside of the home” \cite{BTS24}. Despite the large population in metropolitan city centers, many older transit systems were constructed before accessibility standards were widely adopted, leaving many people with disabilities unable to depend on this form of transit.  Within the 18.6 million disabled Americans who have a travel-limiting disability, many reported reduced travel frequency and an increased dependence on ride-share services as an alternative means of transportation \cite{BTS24}.  

Network science has become a widely used framework for analyzing transportation systems, allowing researchers to characterize structural properties such as connectivity, efficiency, and robustness. Graph-theoretic approaches have been applied extensively to metro systems \cite{Cats17, Derrible09, Derrible11, Ferber07, Ferber09, Latora02, Sien05, Sen03}, bus networks \cite{Huapu07, Sien05, Xu07}, and other infrastructure networks, revealing common structural features such as small-world behavior and heterogeneous degree distributions. These methods have also been used to evaluate node importance, system resilience, and network evolution \cite{Chen23,Chopra16, Daganzo10, Dai22}, as well as to study broader transportation and infrastructure systems including road \cite{Jiang04, Xie07}, air \cite{Guimera05, Li04}, and maritime networks \cite{Ducruet17, Lau24}. Collectively, this body of work demonstrates the utility of network-based approaches for understanding the structure and function of transportation systems.  Parallel to these developments, a substantial literature has emerged focused specifically on accessibility as a multidimensional concept. Foundational frameworks define accessibility in terms of the interaction between transportation systems, land use, and individual needs \cite{Geurs04, Handy20, Levison20, Wu20}. These approaches emphasize that accessibility extends beyond physical infrastructure to include considerations of equity, service provision, and the distribution of opportunities across populations \cite{Bocarejo12, Neutens10, Vanbuckle09}. More recent work has also addressed methodological challenges in measuring accessibility using transit data, including schedule-based and real-time approaches \cite{Wessel19, Liu22}. Together, these studies highlight both the importance and complexity of accessibility in urban systems.

Recent research has increasingly sought to integrate network science with accessibility analysis. Studies have combined graph-theoretic measures with spatial and temporal data to examine connectivity and accessibility within transit networks \cite{Chen14, Luo19}, developed network-based accessibility metrics such as betweenness-accessibility \cite{Sarlas20}, and explored structural similarity and accessibility dynamics using graph-based methods \cite{Jaegal19}. Other work has applied graph theory and GIS techniques to assess accessibility in transportation networks at different spatial scales \cite{Sarker21, Zhou15}. These approaches demonstrate the potential of network-based methods to provide insight into accessibility patterns and their relationship to system structure.  Despite these advances, much of the existing literature focuses either on the structure of the full transit network or on aggregate accessibility measures, without explicitly isolating the subnetwork of stations that meet accessibility criteria. As a result, comparatively less attention has been given to how the structure of accessibility-restricted subnetworks differs from that of the complete system, and therefore, which connections within the system remain available to individuals who rely on step-free access. In addition, while prior studies have explored relationships between accessibility and socioeconomic factors, these analyses are often conducted within a single system or rely on aggregate indices rather than network-based representations that capture structural position within the system. 

This study addresses these gaps by examining infrastructure-based accessibility through a network-theoretic lens. We construct accessibility-restricted subnetworks for two major public transit systems, the London Underground (the Tube) and the New York City Subway (the Metro), where nodes represent accessible stations and edges represent connectivity between them. London and New York City were selected as comparative case studies because both are global metropolitan centers with large, mature rail transit systems that developed over long historical periods, yet they differ substantially in governance, historical development, and the extent and spatial distribution of accessible infrastructure. Much of London's accessible network was built step-free from the outset, including the Jubilee line extension, the Docklands Light Railway (DLR), and the Elizabeth line \cite{TFLnew,TFLel}, while upgrades to existing Underground stations proceed toward a Mayoral target of making half of all Tube stations step-free by 2030, with stations prioritized according to deliverability and available funding \cite{TFLsfa24}. In New York City, by contrast, accessibility has been pursued largely by retrofitting existing stations under a legal mandate. State law required $100$ designated key stations to be made accessible by $2020$, and a $2022$ class-action settlement commits the Metropolitan Transportation Authority (MTA) to adding accessibility at $95\%$ of the remaining inaccessible stations by 2055 \cite{MTAada23}. Engineering constraints have shaped this program directly, as stations judged infeasible to retrofit were replaced on the key-station list by feasible alternates \cite{MTAada23}.  Both systems also provide publicly available data on station accessibility and network structure, enabling consistent comparative analysis. Examining these cities together allows us to compare how similarly sized legacy transit systems with different modernization trajectories have incorporated accessibility improvements. Because two systems cannot by themselves establish general rules for how agencies allocate accessible infrastructure, we treat London and New York City as case studies and introduce a test, applicable to any transit network with station-level accessibility data, that compares the observed placement of accessible stations with random placements of the same number of stations. 

We compare the structural properties of these subnetworks with those of the corresponding full networks, using centrality measures to identify key accessible hubs and the cyclomatic number to measure how many of the system's alternative routes remain available to riders who rely on step-free stations. The contribution of this study is not the introduction of new centrality measures, but rather their application to accessibility-restricted transit subnetworks alongside the full network. By combining comparisons of network structure and station centrality with a random-allocation framework, we examine not only where accessible stations are located within a system, but also whether their structural positions differ from those expected under random placement.  By analyzing these two systems in parallel, we also provide a comparative perspective on how accessibility is distributed across large metropolitan transit networks.  We further examine how accessibility relates to socioeconomic and spatial factors. In the United States, the Americans with Disabilities Act (ADA) mandates that transit agencies identify and update ``key stations” to improve accessibility, which prioritizes high-traffic locations \cite{ADA}.  We test whether accessible stations occupy more central positions in the full network than would be expected if the same number of stations were chosen at random. We also describe how accessible stations are distributed with respect to local median income, daytime population, and ridership. By comparing these patterns across the London and New York transit networks, and relating them to centrality measures in both network types, we aim to better characterize the socioeconomic and structural context in which accessible stations are located.

\section*{Methods}\label{sec:methods}
\subsection*{Definition of accessibility}\label{sec:doa}
Throughout this paper, ``accessible'' is used in a narrow, infrastructural sense. A station is classified as accessible if the transit agency designates it as step-free, meaning a passenger using a wheelchair can travel between the street and the platform by elevator, ramp, or level route without stairs or escalators. This differs from the broader use of the term in the transport literature, where accessibility refers to the ease of reaching opportunities such as employment and services through the combined transport and land-use system \cite{Geurs04, Levison20}. Our analysis concerns only the former.

For London, accessible stations were identified from the step-free symbols on the Transport for London Tube map, which distinguishes stations offering step-free access from street to train from those offering step-free access from street to platform only, where a step or gap remains at boarding. Both categories were included in the accessible network. In the network model, every accessible station is treated as providing full step-free access from street to train and between the accessible lines that serve it, with no gap at boarding and all elevators in working order. This is an idealization. Stations that are step-free only to the platform, platform gaps, and elevator outages all reduce usability in practice \cite{NYCaccsize}, so our results describe an upper bound on the connectivity available to riders who rely on step-free infrastructure. For New York City, accessible stations were identified from the MTA Accessible Stations Subway Map. In both cities, stations accessible in only one direction were excluded, and where a station is accessible on only some of the lines serving it, only those lines contribute edges to the accessible network. Although both systems are evaluated using agency-designated step-free access, the underlying accessibility criteria and station classifications are not necessarily identical; therefore, cross-city comparisons should be interpreted as comparisons of agency-designated infrastructure accessibility rather than as measurements under a single harmonized accessibility standard.

Agency station counts differ from the number of nodes in our networks. In London, the totals of 92 Tube, 41 Elizabeth line, and 45 DLR accessible stations overlap, because interchange stations such as Stratford, Canary Wharf, and Whitechapel are counted in each system that serves them. Each such station is a single node in our network, giving 162 accessible nodes. In New York City, stations within a single complex, such as Times Square--42nd Street, are counted separately by the MTA but represented as one node, giving 125 accessible nodes from the 130 accessible stations.

\subsection*{Data}\label{sec:data}
Transport for London is actively working to convert inaccessible stations into fully accessible ones. As of $2025$, the Tube spans approximately $250$ route miles of track, which includes $272$ stations, of which $92$ are accessible.  In addition, all $41$ stations on the Elizabeth line and all $45$ stations on the DLR meet accessibility standards \cite{TFLdatatubesize,TFLel}. Although newly constructed stations and lines are designed to be accessible from their conception, some existing stations offer only step-free access from street to platform, leaving significant challenges for boarding due to gaps or height differences between the platform and the train. While notable progress has been made in improving accessibility across London’s transit infrastructure, substantial work remains before the system becomes fully accessible \cite{TFLel}. For the Tube, data describing the full network was originally compiled by Nicola Greco \cite{LonTota}, but this dataset required updates to incorporate new lines and stations. The accessible network was constructed by combining the updated full-network data with the London Underground step-free access map, ``Step-Free Tube Guide Map". London’s transport system is organized into numbered zones, which we use to characterize the spatial distribution of stations: Zone 1 is the most central, while outer zones extend farther from central London.

New York’s Metropolitan Transportation Authority is similarly working to increase accessibility for riders with ambulatory disabilities. At the time our network data was compiled, only $28\%$ of the city’s $472$ subway stations, which span approximately $248$ route miles of track, were considered accessible, totaling $130$ stations \cite{NYCdatametrosize, NYCaccsize}. Moreover, audits have shown that many elevators within these stations are improperly labeled, poorly maintained, or frequently out of service, further limiting effective accessibility for transit riders\cite{NYCaccsize}. For the NYC subway, station names and line information were obtained from the NYC Open Data database \cite{NYCTota}. The accessible network was constructed using this full-network data along with the MTA’s ``Accessible Stations Subway Map".

All network analyses for both cities were conducted in Python using the NetworkX package. 

\subsubsection*{Socioeconomic and spatial data}
To examine the broader urban context of accessible stations, station latitude and longitude coordinates were linked to administrative geographic units using geospatial boundary files and spatial joins conducted in Python with the GeoPandas package.

Station names and geographic coordinates for London Underground stations were obtained from the updated Greco dataset \cite{LonTota} and were spatially joined to London borough polygons obtained from a publicly available GeoJSON boundary dataset. Borough-level median income and daytime population data were obtained from the London Datastore and linked to stations based on borough membership.

Station names, coordinates, and line information for New York City subway stations were obtained from NYC Open Data \cite{NYCTota} and were spatially joined to official NYC Community District boundary files obtained through NYC Open Data. Median income data were obtained from the U.S. Census Bureau’s American Community Survey (ACS) 5-year estimates (2021), using table B19013 (Median Household Income in the Past 12 Months), accessed through the Census API. Tract-level estimates were aggregated to community districts by taking the median across the tracts in each district.  District-level income values were linked to stations according to their assigned community district. Station-level weekday and weekend ridership were obtained from Metropolitan Transportation Authority ridership reports and used descriptively as indicators of commuter and visitor concentration.

The geographic units used in the two cities are not directly equivalent. London boroughs are larger municipal administrative areas, whereas New York City community districts are smaller planning units nested within boroughs. These units were selected because they provide publicly available and consistently reported socioeconomic data within each city. Accordingly, comparisons involving these variables should be interpreted as within-city spatial patterns rather than strict cross-city equivalence of administrative areas.  While finer spatial units may offer greater local precision, borough- and district-level data provide a practical basis for comparative analysis.

\subsection*{Model}\label{sec:model}
In order to examine the topological structure of our transit networks, we represent them as a graph, $G(V,E)$, where $V$ is the set of nodes and $E$ is the set of edges, and the number of nodes and edges is given by $N$ and $M$, respectively. The stations in the transit system are represented as nodes and edges exist between them if these stations are adjacent on the line.  In the accessible networks, nodes represent accessible stations, and edges indicate connectivity between them. Unlike in the full network, where edges correspond to direct adjacency along a line, accessible stations are often not consecutive stops. Consequently, in constructing the accessible networks, we define an edge between two accessible stations whenever one is the next accessible stop along a given line, regardless of how many inaccessible stations lie between them in the complete transit network. For example, Pelham Bay Park is connected to Westchester Square as they are the two adjacent accessible stations on the $6$ line. This type of network topology, in terms of public transit networks, is called $\mathbb{L}$-space \cite{Sen03,Sien05}.

This network is undirected, since transit lines operate in both directions, and unweighted. Because many lines overlap within the system, a single station may be served by multiple transit lines. For example, stations along the Circle line may also be served by the District line or the Hammersmith \& City line. To avoid the introduction of multi-edges, any duplicated connections are represented by a single edge. 

Most accessible stations also support accessible transfers between lines. For stations served by multiple lines where only a subset is accessible, these distinctions were recorded. Stations that share a name but correspond to different physical locations or lines are clarified using parenthetical labels. For example, the two stations named Gun Hill Road, located on the $2$ and $5$ lines respectively, are distinguished accordingly.

\subsection*{Degree distribution}\label{dd}
Previous works find that many infrastructure networks modeled in $\mathbb{L}$-space are scale-free\cite{Guimera05, Li04}, including metro \cite{Derrible11, Sen03, Sien05, Ferber07, Ferber09}, bus \cite{Huapu07, Luo19, Sien05, Xu07}, and road \cite{Xie07} networks.  Motivated by these findings, we explore the degree distribution of our networks and determine whether our networks also display scale-free behavior.  The indication that a network is scale-free is if the degree distribution of the network is power-law, i.e. 
\begin{eqnarray}
    p(k)\sim k^{-\gamma},
\end{eqnarray}
where $\gamma$ is a constant and $k$ is the degree. Values for this constant are typically in the range of $2\leq \gamma \leq 3$ \cite{Newman}.

\subsection*{Closeness centrality}\label{cc}
Closeness centrality measures the mean distance from node $i$ to the other nodes in the network. Then the mean shortest distance from $i$ to all others is 
\begin{eqnarray}\ell_i=\frac{1}{N-1}\sum_j d_{ij},\end{eqnarray}
where $d_{ij}$ is the shortest distance from node $i$ to node $j$, and $N$ is the number of nodes. Mean distance is normally not considered a centrality measure because it produces results that are opposite of what is expected.  More central nodes have a smaller mean shortest distance, and vice versa \cite{Newman}. Thus, the inverse of the mean shortest distance is taken to obtain the centrality measure called closeness centrality. Thus, the closeness centrality of node $i$ is given by 
\begin{align}
    c_i&=\frac{1}{\ell_i}\notag\\
    &=\frac{N-1}{\sum_{j=1}^N d_{ij}}
\end{align}
Notice that a low value for $\ell_i$ would create a high centrality score, $c_i$.  Likewise, a high value for $\ell_i$ would create a low centrality score, $c_i$. Ultimately, closeness centrality calculates the shortest paths between all nodes, then assigns each node a score based on the sum of its shortest paths. A node with a closeness centrality of $0$ is completely isolated, and a node with a closeness centrality score of $1$ is connected to all other nodes.

In our specific network, a node with high closeness centrality is more central than one with low closeness centrality. In a highly connected network, the nodes in the network obtain closeness centrality scores that are close in value.  Also, it has been found that nodes with higher degree have shorter average distance to others, so closeness centrality and degree are positively correlated \cite{Newman}. Because an edge in the accessible network may span several inaccessible stations, and the accessible network contains fewer nodes than the full network, raw closeness values in the accessible network are not directly comparable to those in the full network of the same city. We therefore compare closeness across the accessible and full networks only through rank correlations.

\subsection*{Betweenness centrality}\label{bc}
Betweenness centrality measures how often a node lies on the shortest path between other nodes.  Let $n_{st}^i$ be the number of shortest paths between $s$ and $t$ that pass through node $i$, and let $g_{st}$ be the total number of shortest paths between $s$ and $t$. The betweenness centrality of node $i$ is
\begin{eqnarray}
x_i=\frac{2}{(N-1)(N-2)}\sum_{s<t,\; s,t\neq i} \frac{n_{st}^i}{g_{st}}.
\end{eqnarray}
If every pair of stations exchanged equal traffic along shortest paths, this quantity would be proportional to the rate at which traffic passes through node $i$ \cite{Newman}. Because our networks contain no ridership or service information, we interpret betweenness structurally. A station with high betweenness lies on many shortest paths between other stations, and in the accessible network such a station is a key transfer point for step-free routes. 

Because $x_i$ is normalized by the number of pairs of other stations, it gives the share of those pairs whose shortest paths pass through $i$, counting a pair fractionally when only some of its shortest paths do. We therefore compare betweenness values between networks as shares of station pairs, although the pairs and the shortest paths between them differ between the accessible and full networks.

\subsection*{Cyclomatic number}\label{cn}
To compare the overall structure of the accessible and full networks, we use the cyclomatic number, a classical measure of transport network structure \cite{Kansky63, Derrible11}. For a graph with $N$ nodes, $M$ edges, and $p$ connected components, the cyclomatic number is
\begin{eqnarray}
\mu = M - N + p,
\end{eqnarray}
which counts the number of independent cycles in the network. In a transit network, each independent cycle provides an alternative route between stations, so $\mu$ measures the redundancy of the network \cite{Derrible11}. A network with $\mu = 0$ is a tree, with exactly one path between any two stations. All four networks in this study are connected, so $p = 1$.

We use $\mu$ rather than path-based measures such as the network diameter to compare the accessible and full networks. Because an edge in the accessible network may span several inaccessible stations, path lengths in the two networks are not measured in the same units. The cyclomatic number does not have this problem, since replacing a run of inaccessible stations with a single edge removes the same number of nodes as edges and leaves $\mu$ unchanged. A decrease in $\mu$ from the full network to the accessible network therefore generally reflects alternative routes that are no longer available to riders who rely on accessible stations. One exception arises when two routes between the same pair of accessible stations, such as an express and a local service, pass only through inaccessible stations. In the accessible network these routes become a single edge, so the cycle they formed is removed even though both routes remain usable. Because express and local services run in parallel more often in New York City than in London, the difference in $\mu$ between the two cities may partly reflect this effect.

\subsection*{Rank correlation of centralities}\label{rc}
To compare centrality rankings across measures and across networks without relying on a fixed cutoff, we use Spearman's rank correlation coefficient \cite{Spearman04}, a standard tool for comparing node centralities \cite{Oldham19}. For two quantities, $x$ and $y$, measured on the same $n$ stations, each is converted to ranks, $r_x$ and $r_y$, with tied values assigned the average of their ranks, and
\begin{eqnarray}
\rho = \frac{\mathrm{cov}(r_x, r_y)}{\sigma_{r_x}\,\sigma_{r_y}},
\end{eqnarray}
where $\mathrm{cov}(r_x, r_y)$ is the covariance of the two sets of ranks and $\sigma_{r_x}$ and $\sigma_{r_y}$ are their standard deviations, so that $\rho$ is the Pearson correlation of the ranks. The coefficient $\rho$ ranges from $-1$ to $1$ and measures how consistently the two quantities, $x$ and $y$, order the stations, whether or not the relationship between them is linear.

We compute $\rho$ in two settings. Within each accessible network, we correlate the betweenness and closeness centrality of every station. For each accessible station, we also correlate its centrality in the accessible network with its centrality in the full network, separately for betweenness and closeness. For this second comparison, station names that differ between the accessible and full network datasets were matched manually.

\subsection*{Comparison with random allocation}\label{ra}
To assess whether the placement of accessible stations differs from what would be expected by chance, we compare the set of accessible stations $A\subset V$ with randomly chosen sets of the same size in the full network $G$. Comparing an observed set of nodes with random sets of the same size, optionally preserving node degree, is an established way to test whether a set of nodes occupies a distinctive position in a network \cite{Menche15,Guney16}, and random selection of stations is the standard baseline in studies of transit network resilience \cite{Berche09,vonFerber12}. All quantities in this section are computed on the full network, so that the observed and random sets are evaluated on the same graph.

We summarize each set of stations by the mean full-network betweenness centrality, closeness centrality, and degree of its members, and by two coverage measures. In the main text we report mean betweenness centrality and the coverage measure $C_2$ defined below, and results for all statistics are given in Appendix~\ref{app:rat}. Let 
\begin{equation}
d(i,A)=\min_{a\in A} d_{ia}
\end{equation} 
be the shortest-path distance, in stops, from station $i$ to the nearest accessible station. The mean distance from an inaccessible station to the nearest accessible station is
\begin{eqnarray}
H(A)=\frac{1}{|V\setminus A|}\sum_{i\in V\setminus A} d(i,A),
\end{eqnarray}
and the share of stations within two stops of an accessible station is
\begin{eqnarray}
C_2(A)=\frac{1}{N}\left|\{\,i\in V \mid d(i,A)\le 2\,\}\right|.
\end{eqnarray}
The two-stop threshold follows the coverage goal set by the MTA \cite{MTAada23}. Because distances are measured in stops, these measures describe position within the network rather than geographic distance.

We generate $R=1000$ random sets $A'$ with $|A'|=|A|$ under two null models. In the uniform null model, $A'$ is drawn uniformly at random from $V$. In the degree-matched null model, the number of stations of degree $1$, $2$, $3$, and $\geq 4$ in $A'$ equals that in $A$, which controls for the tendency of interchange stations to be selected. For each statistic $s$, we report the percentile of the observed value $s(A)$ within the null distribution and the two-sided empirical $p$-value
\begin{eqnarray}
p=\min\left\{1,\;2\min\left(\frac{1+\#\{s(A')\geq s(A)\}}{R+1},\;\frac{1+\#\{s(A')\leq s(A)\}}{R+1}\right)\right\}.
\end{eqnarray}
This estimator includes the observed set in the reference distribution and therefore never returns a $p$-value of zero \cite{North02}. Because the accessible networks were compiled manually from the agency step-free maps, reconstructing an $\mathbb{L}$-space accessible network for each of the $R$ random sets is not practical, so the random sets are compared using full-network statistics.

\section{Results}\label{sec:results}

\subsection{London Tube Network}
\subsubsection{Accessible Network}
The accessible network of the London Underground has $162$ nodes and $195$ edges, giving a cyclomatic number of $\mu = 34$. As shown in Table \ref{tab:lona}(a), Stratford exhibits the highest betweenness centrality within our network, with a value of $0.488$.  Stratford lies on the Central line and provides extensive interchange opportunities, connecting to the Jubilee, Elizabeth, and DLR lines.  Of these four lines, the Elizabeth and DLR are completely accessible and the Jubilee is $63\%$ accessible. This value means that roughly half of the shortest step-free paths between other pairs of accessible stations pass through Stratford. As a result, Stratford is the most structurally central interchange in the accessible network.

\begin{table}[!t]
\caption{The ten nodes in the London Accessible network with the highest centrality measures.}
\vspace{-10pt}
\begin{adjustwidth}{-.05in}{.25in}
\begin{subtable}[h]{.4\textwidth}
\centering
\caption{Betweenness}
\center
\def\arraystretch{1.25}
\begin{tabular}{|c|c|} 
\hline
Node Name & Betweenness\\
\hline\hline
Stratford&0.488\\\hline
Tottenham Court Road&0.415\\\hline
Bond Street&0.321\\\hline
Canary Wharf& 0.256\\\hline
Whitechapel&0.236\\\hline
Ealing Broadway& 0.219\\\hline
King's Cross St. Pancras& 0.214\\\hline
West Ham & 0.168\\\hline
Paddington & 0.137\\\hline
West Ealing & 0.137\\
\hline
\end{tabular}
\end{subtable}%
\hfill
\begin{subtable}[h]{0.4\textwidth}
\centering
\caption{Closeness}
\center
\def\arraystretch{1.25}
\begin{tabular}{|c|c|}
\hline
Node Name &  Closeness\\
\hline\hline
Tottenham Court Road&0.199\\\hline
Stratford& 0.196\\\hline
Bond Street&0.191\\\hline
Farringdon& 0.185\\\hline
Whitechapel& 0.183\\\hline
Finchley Central& 0.179\\\hline
King's Cross St. Pancras & 0.178\\\hline
Golders Green&0.178\\\hline
Liverpool Street& 0.177\\\hline
Paddington& 0.174\\
\hline
\end{tabular}
\end{subtable}
\begin{flushleft} 
rounded to the third decimal place
\end{flushleft}
\label{tab:lona}
\end{adjustwidth}
\end{table}

\begin{figure}[b!]
\centering
\includegraphics[width=170mm]{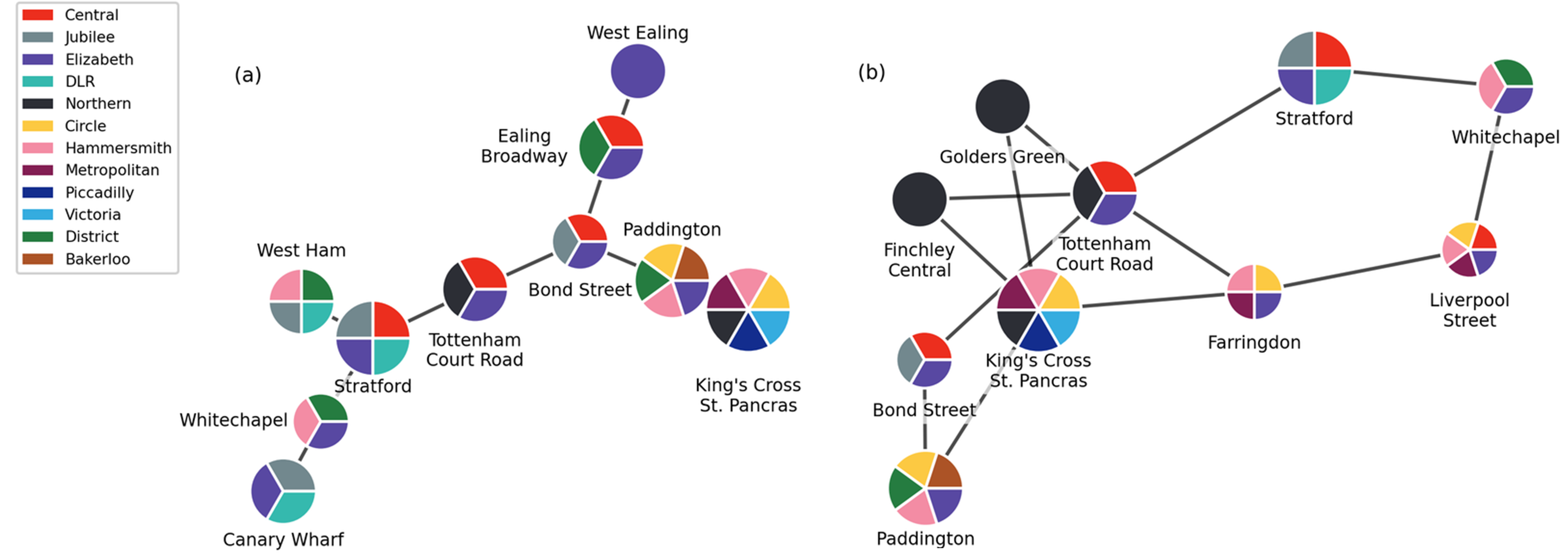}
\caption{Subgraphs of the ten nodes with the highest (a) betweenness and (b) closeness centralities.  Colored pieces of the nodes represent the lines that stop at them and size of the node is based on the degree of the node in the whole accessible network.}
\label{fig:lonasubgraph}
\end{figure}

Fig.~\ref{fig:lonasubgraph}(a) illustrates that four of the ten most central nodes are on the Central line; however, eight of the ten are on the Elizabeth line, whose route is broadly aligned with the Central line but is fully accessible. Together, these observations indicate that the central stations play a key structural role in connecting the accessible network.

Conversely, King's Cross St. Pancras does not lie on the Central or Elizabeth line, yet it remains one of the most structurally important stations in the network. It serves six Underground lines: the Piccadilly, Victoria, Northern, Hammersmith \& City, Circle, and Metropolitan lines, providing extensive interchange opportunities across the system. In addition, King’s Cross St. Pancras is a major hub for the National Rail service and lies on the Piccadilly line, which connects directly to Heathrow Airport terminals. These factors collectively underscore why King’s Cross St. Pancras functions as a critical node within the accessible network.

Turning to Table \ref{tab:lona}(b), we observe that Tottenham Court Road has the highest closeness centrality, with a value of $0.199$, indicating that it can reach a large share of accessible stations through relatively short paths. Because it also has the second-highest betweenness centrality, Tottenham Court Road is both close to the rest of the accessible network and a major transfer point within it.

Across all $162$ stations of the accessible network, betweenness and closeness rankings are only weakly correlated (Spearman's $\rho = 0.28$, $p < 0.001$) \cite{Spearman04, Oldham19}, so the stations that lie on the most step-free shortest paths are often not those closest. Six stations, King's Cross St. Pancras, Whitechapel, Bond Street, Tottenham Court Road, Stratford, and Paddington, appear among the ten highest in both measures. The closeness centrality values are relatively similar across the top-ranked stations, but, notably, all fall below $0.2$. This indicates that even the most central stations are not close to the majority of nodes in the accessible network. In other words, they are not centrally located with respect to the full structure of the network.  Finchley Central and Golders Green rank highly because the Northern line stations between them and central London are inaccessible, so a single edge in the accessible network spans that entire run.

This observation is consistent with the geography of the Underground system. Many lines extend into outer zones (i.e. Zones $3–6$ and beyond), where accessible stations frequently have degree $1$ or $2$, and where lines tend not to interconnect. As in a spiderweb, the farther one moves from the core of the system, the sparser the connectivity becomes. The presence of long accessible corridors, such as the Elizabeth line and the DLR, further contributes to larger average distances between stations. These structural features explain why closeness centrality scores across the network remain comparatively low.

\begin{table}[!b]
\caption{The ten nodes in the London Tube network with the highest centrality measures and whether they are accessible.}
\vspace{-10pt}
\begin{adjustwidth}{.05in}{.1in}
\begin{subtable}[h]{.4\textwidth}
\centering
\caption{Betweenness}
\center
\def\arraystretch{1.25}
\hspace*{-20mm}
\begin{tabular}{|c|c|c|}
\hline
Node Name & Accessible?&  Betweenness\\
\hline\hline
Bond Street &Y & 0.450\\\hline
Liverpool Street &Y& 0.429\\\hline
Farringdon &Y & 0.402\\\hline
Whitechapel &Y & 0.398\\\hline
Tottenham Court Road &Y& 0.322\\\hline
Paddington &Y & 0.293\\\hline
Stratford &Y& 0.291\\\hline
Baker Street&N& 0.289\\\hline
Ealing Broadway &Y& 0.204\\\hline
Finchley Road &N& 0.199\\
\hline
\end{tabular}
\end{subtable}%
\hfill
\begin{subtable}[h]{0.4\textwidth}
\centering
\caption{Closeness}
\center
\def\arraystretch{1.25}
\hspace*{-10mm}
\begin{tabular}{|c|c|c|}
\hline
Node Name & Accessible?&  Closeness\\
\hline\hline
Bond Street &Y& 0.131\\\hline
Farringdon &Y& 0.130\\\hline
Tottenham Court Road &Y& 0.130\\\hline
Liverpool Street &Y& 0.127\\\hline
Oxford Circus&N& 0.126\\\hline
Paddington &Y&0.122\\\hline
Baker Street&N& 0.122\\\hline
Green Park &Y& 0.122\\\hline
King's Cross St. Pancras &Y& 0.121\\\hline
Whitechapel &Y& 0.120\\
\hline
\end{tabular}
\end{subtable}
\begin{flushleft} rounded to the third decimal place
\end{flushleft}
\label{tab:lontot}
\end{adjustwidth}
\end{table}

\subsubsection{Full Network}
The full London Underground (including the DLR and Elizabeth lines, but excluding the Overground lines) network has $337$ nodes and $397$ edges, giving a cyclomatic number of $\mu = 61$. The accessible network therefore retains $34$ of the $61$ independent cycles of the full network ($56\%$), so just over half of the alternative routes in the system remain available to riders who rely on accessible stations. We now compare the centrality measures of the accessible network with those of the complete network. This comparison allows us to examine whether the top-ranked stations in the accessible network occupy similarly central positions in the full system, and to determine how many of the most central stations in the complete network are themselves accessible.

\begin{figure}[b!]
\centering
\includegraphics[width=170mm]{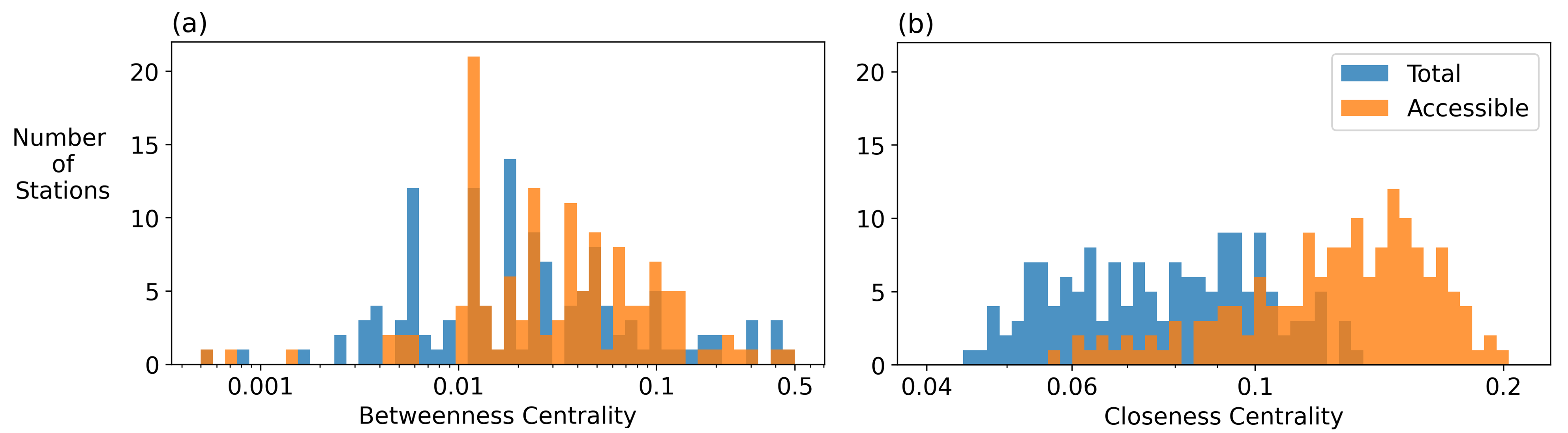}
\caption{Number of stations in the full and accessible networks by (a) betweenness and (b) closeness centrality. Raw values are not directly comparable between the two networks.}
\label{fig:loncomall}
\end{figure}

As shown in Table \ref{tab:lontot}(a), eight out of our top ten stations for the total network are accessible.  Bond Street attains the highest betweenness centrality in the complete network, with a value of $0.450$, which is substantially larger than its betweenness value in the accessible network, which is $0.321$. This increase is attributable to the structure of the lines serving Bond Street, namely the Central, Elizabeth, and Jubilee lines, and in particular to its numerous connections to non-accessible stations along these routes. 

The centrality measures of the full London network reflect station position within the entire transit system rather than only among accessible stations. Stations with high betweenness centrality, such as Bond Street and Liverpool Street, function as major transfer points that lie on many shortest paths in the full network. Their prominence in both the full and accessible networks shows that several of London's most important interchange stations are accessible. Whether this reflects deliberate prioritization is examined below by comparison with random allocations. However, the presence of non-accessible stations such as Baker Street and Finchley Road among the most central nodes of the full network shows that some structurally important parts of the broader system remain less accessible. This contrast suggests that while London has made progress in upgrading major hubs, accessibility coverage is still incomplete at certain high-value network locations.

\begin{figure}[t!]
\centering
\includegraphics[width=170mm]{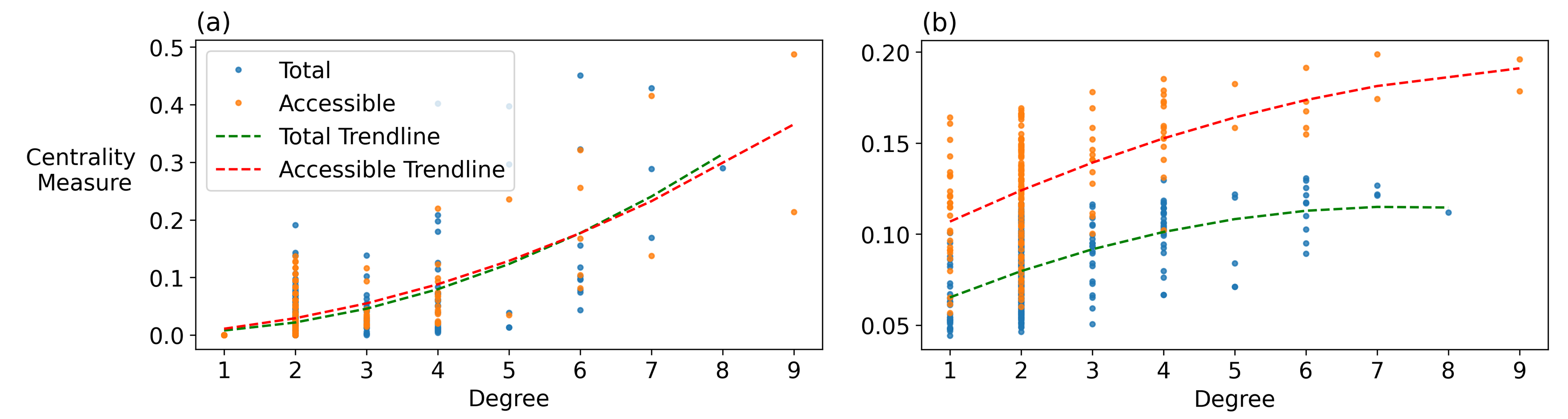}
\caption{(a) Betweenness and (b) closeness centrality values compared to degree of the nodes for the London Underground network.  Trendlines are added to visualize correlation between these centrality measures and degree. Raw values are not directly comparable between the two networks.}
\label{fig:loncomgrap}
\end{figure}

Panel (a) of Fig.~\ref{fig:loncomall} shows that many stations in both networks exhibit low betweenness centrality, with most of the values below $0.1$, while a small number of stations, listed in Tables \ref{tab:lona} and \ref{tab:lontot}, reach much higher values. Panel (b) shows the corresponding distributions of closeness centrality. As noted in the Methods, raw closeness values are not comparable between the accessible and full networks, so we compare closeness across the two networks through the rank correlations reported below.

These results provide insight into the distribution of accessible stations and the correspondence between the accessible network and the full network. In the total network, eight of the ten stations with the highest betweenness centrality are accessible, indicating that many of the most structurally important nodes are accessible even when considering all stations. Comparing each accessible station's centrality in the two networks shows that restricting the system to accessible stations largely preserves their relative importance. The rank correlation between a station's betweenness in the accessible and full networks is $\rho = 0.86$, and for closeness it is $\rho = 0.75$ ($p < 0.001$ in both cases, $n = 162$). Overall, these stations are structurally central in both the accessible and full networks.

Next, we examine the relationship between node degree and centrality measures. As shown in Fig.~\ref{fig:loncomgrap}(a), there is a positive correlation between betweenness centrality and node degree, a trend that holds in both the accessible and full networks. Closeness centrality also exhibits a positive correlation with degree in both networks; however, in the full network, this correlation diminishes for nodes with higher degree, as illustrated in Fig.~\ref{fig:loncomgrap}(b). 

\begin{figure}[!t]
\centering
\includegraphics[width=80mm]{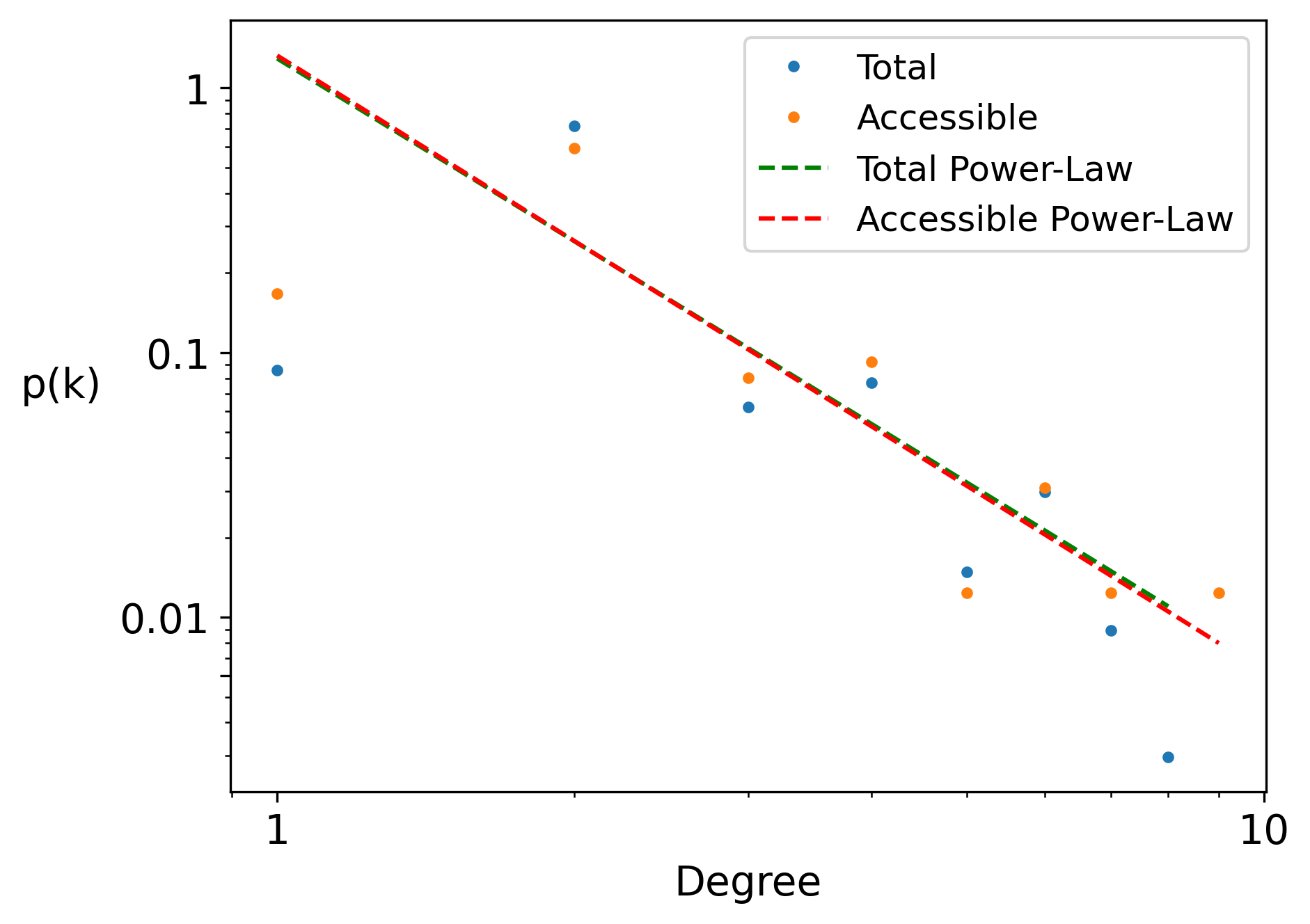}
\caption{Degree distribution of the London accessible and full networks. Shown on logarithmic axes with power-law fits}
\label{fig:londonpl}
\end{figure}

Lastly, we investigate whether our networks exhibit scale-free behavior, characterized by a power-law degree distribution.  As shown in Fig.~\ref{fig:londonpl}, the degree distribution of both networks approximately follow a power law, with exponents $\gamma_{A1}=2.327$ for the accessible network and $\gamma_{T1}=2.293$ for the full network. Because the degrees in both networks span less than one order of magnitude, we cannot determine with certainty that these distributions follow a power law \cite{Clauset09}, but both display power-law-like behavior with exponents in the range typical of scale-free networks. These results are consistent with previous studies indicating that public transit networks in $\mathbb{L}$-space display scale-free properties \cite{Derrible11, Sen03, Sien05, Ferber07, Ferber09}.

\begin{figure}[!b]
\centering
\includegraphics[width=100mm]{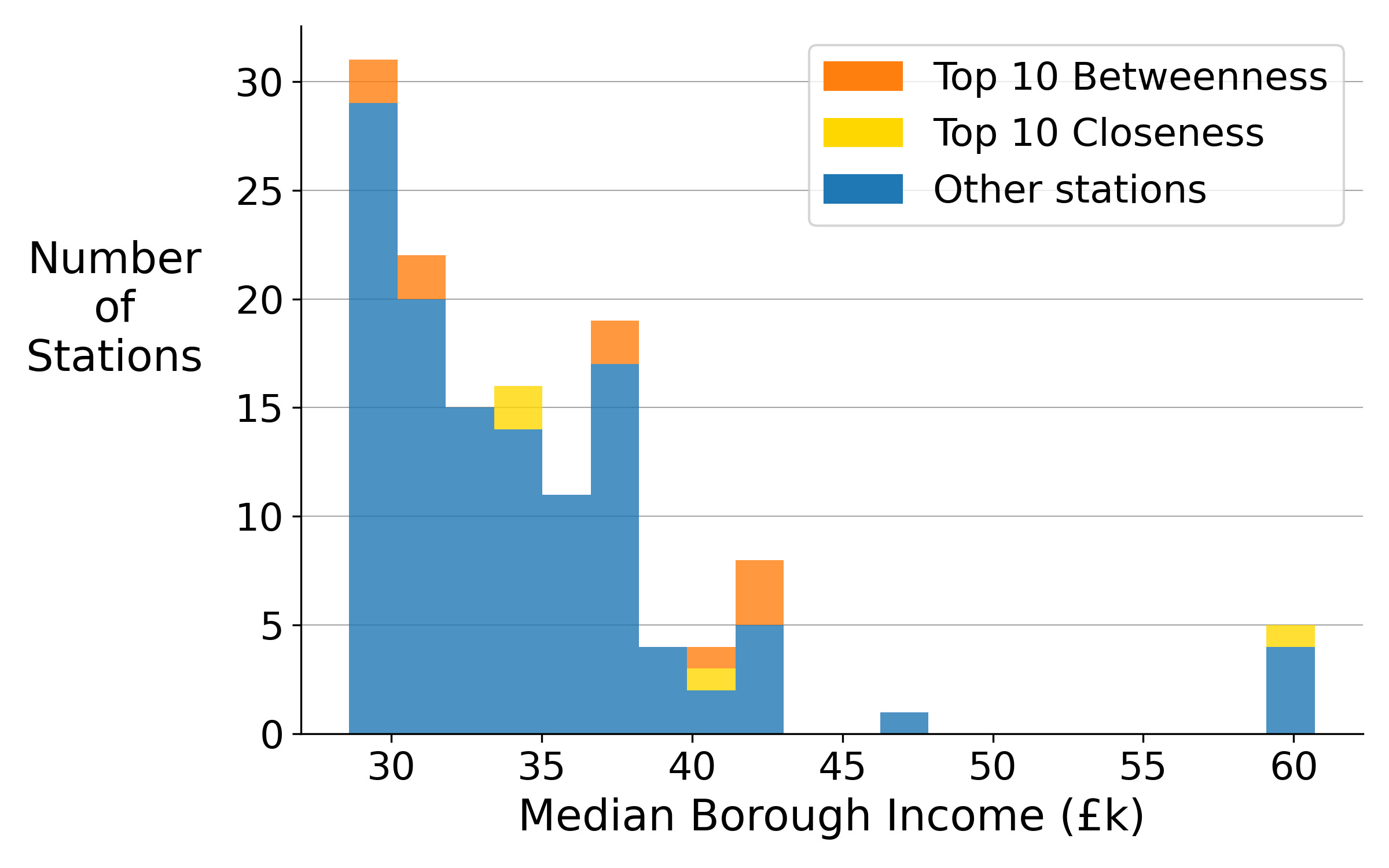}
\caption{Median annual income, per thousand pounds (\textsterling), per borough and the number of accessible stations in each borough. The accessible stations in our top ten list of betweenness and closeness centralities and their median annual income are indicated in orange and yellow, respectively.}
\label{fig:lonincome}
\end{figure}

\subsubsection{Income and Tourism Results}
As shown in Fig. \ref{fig:lonincome}, accessible stations are distributed across boroughs with a wide range of median income levels, and the relationship between borough income and accessibility does not appear strictly monotonic. Many accessible stations are located in boroughs with moderate income levels, while some higher-income central boroughs also contain important accessible stations. For example, Newham borough contains approximately $23$ accessible stations, out of the $25$ total stations, despite a median income of \textsterling$28.9\text{k}$. Much of this reflects lines that were built step-free, including the $1999$ Jubilee line extension, the DLR, and the Elizabeth line, several of which share interchange stations such as Stratford, West Ham, and Canning Town \cite{TFLnew}. Nevertheless, this suggests that income alone does not explain the spatial distribution of accessibility across London.

\begin{figure}[t!]
\centering
\includegraphics[width=170mm]{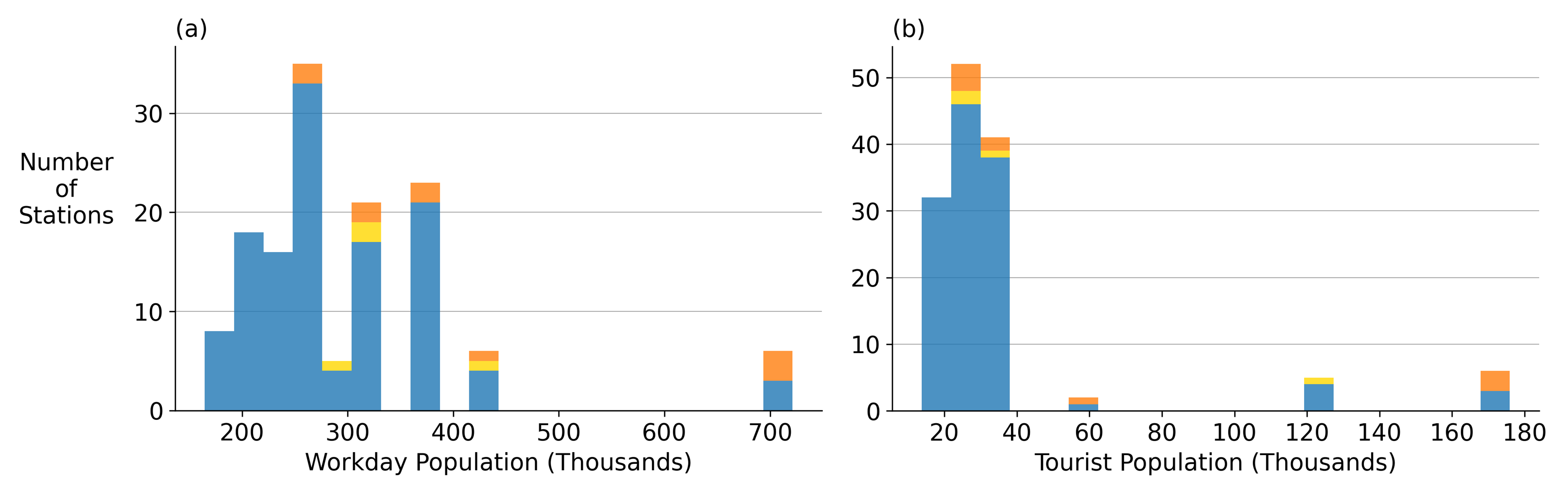}
\caption{(a) Worker and (b) tourist daytime population per borough and the number of accessible stations in each borough. The accessible stations in our top ten list of betweenness and closeness centralities and their daytime population are indicated in orange and yellow, respectively.}
\label{fig:lonpop}
\end{figure}

By contrast, Fig.~\ref{fig:lonpop}(a) shows an association between accessible stations and areas with larger daytime populations, particularly worker concentrations and mixed-use central districts. Many stations with high betweenness and closeness centrality are located in boroughs that experience substantial daytime inflows, consistent with their role as employment centers, interchange locations, and major destinations. This pattern is consistent with accessible stations being concentrated in boroughs with high daytime activity, although borough-level data cannot establish why. Figure \ref{fig:lonpop}(b) shows the corresponding distribution for tourist population. Most accessible stations lie in boroughs with comparatively small tourist populations, while the few boroughs with the largest tourist populations contain several of the top-ranked stations.

Overall, the London results indicate that accessibility is concentrated in several high-activity and highly connected parts of the city, while outer portions of the network often remain more limited in accessible coverage. These patterns may reflect a combination of historical investment decisions, network expansion, engineering feasibility, and passenger demand rather than any single socioeconomic factor.

\begin{table}[!t]
\caption{The ten nodes in the New York Accessible network with the highest centrality measures.}
\vspace{-10pt}
\begin{adjustwidth}{-.25in}{.25in}
\begin{subtable}[h]{.4\textwidth}
\centering
\caption{Betweenness}
\center
\def\arraystretch{1.25}
\hspace*{-10mm}
\begin{tabular}{|c|c|} 
\hline
Node Name & Betweenness\\
\hline\hline
Grand Central - 42nd Street&0.285\\\hline
Times Square - 42nd Street& 0.276\\\hline
Herald Square - 34th Street& 0.256\\\hline
125th Street (4)& 0.232\\\hline
Fulton Street& 0.221\\\hline
Union Square - 14th Street &0.195\\\hline
Atlantic Ave - Barclay's Center&0.185\\\hline
Jay Street - MetroTech& 0.180\\\hline
Brooklyn Bridge - City Hall& 0.161\\\hline
Marcy Ave& 0.136\\
\hline
\end{tabular}
\end{subtable}%
\hfill
\begin{subtable}[h]{0.4\textwidth}
\centering
\caption{Closeness}
\center
\def\arraystretch{1.25}
\hspace*{-11mm}
\begin{tabular}{|c|c|}
\hline
Node Name &  Closeness\\
\hline\hline
Times Square - 42nd Street&0.247\\\hline
Herald Square - 34th Street& 0.246\\\hline
Grand Central - 42nd Street& 0.239\\\hline
Fulton Street& 0.238\\\hline
Brooklyn Bridge - City Hall& 0.232\\\hline
West 4th Street - Washington Square &0.232\\\hline
Union Square - 14th Street& 0.230\\\hline
Broadway - Lafayette Street & 0.221\\\hline
47th-50th Streets - Rockefeller Center &0.220 \\\hline
Jay Street - MetroTech &0.217\\
\hline
\end{tabular}
\end{subtable}
\begin{flushleft} rounded to the third decimal place.
\end{flushleft}
\label{tab:nyca}
\end{adjustwidth}
\end{table}

\subsection{New York City Metro Network}
\subsubsection{Accessible Network}
\begin{figure}[b!]
\centering
\includegraphics[width=170mm]{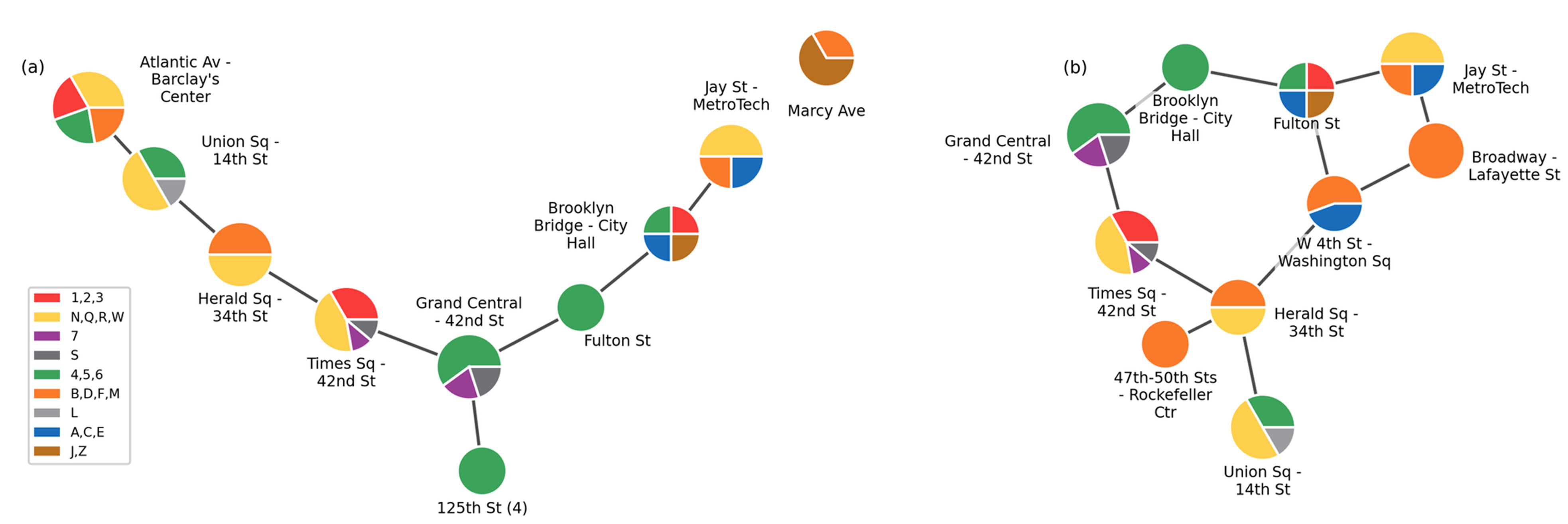}
\caption{Subgraphs of the ten nodes with the highest (a) betweenness and (b) closeness centralities.  Colored pieces of the nodes represent the lines that stop at them and size of the node is based on the degree of the node in the whole accessible network.}
\label{fig:nycasubgraph}
\end{figure}

The accessible network for the New York City Metro has $125$ nodes and $162$ edges, giving a cyclomatic number of $\mu = 38$.  As shown in Table \ref{tab:nyca}(a), Grand Central–$42$nd Street has the highest betweenness centrality in the accessible network, with a value of $0.285$. Although this station serves fewer lines than some other major interchanges, it is directly connected to another major hub, Times Square. As illustrated in Fig. \ref{fig:nycasubgraph}, many of the stations with high betweenness centrality lie on the green ($4,5,6$) and yellow (N,Q,R,W) lines. These stations therefore lie on many of the shortest step-free paths within the accessible network.

It should also be noted that several stations provide connections to additional public transit systems. Grand Central Station, for instance, is not only a major Metro stop but also serves as a transfer point for the Long Island Rail Road and the Metro-North Commuter Railroad. These connections further emphasize its importance as a central node within the network.

Finally, we note that one station in Fig. \ref{fig:nycasubgraph}(a), Marcy Avenue, is not directly connected to any of the other stations in the top ten for betweenness centrality in the accessible network. Its high betweenness comes instead from linking Broadway--Lafayette Street, which has a betweenness centrality of $0.124$, with the accessible stations farther out along its line in Brooklyn and Queens.

Now, by examining Table \ref{tab:nyca}(b), we find that Times Square–42nd Street has the highest closeness centrality, with a value of $0.247$. Across all $125$ stations of the accessible network, betweenness and closeness rankings are moderately correlated (Spearman's $\rho = 0.58$, $p < 0.001$), considerably more strongly than in London ($\rho = 0.28$). This is consistent with a compact accessible core in which the stations that lie on the most step-free shortest paths are also close to the rest of the accessible network. Seven stations appear among the ten highest in both measures.

Many lines extend outward from Manhattan into the Bronx, Queens, and Brooklyn. In these outer boroughs, accessible stations typically have degree $1$ or $2$. Consequently, the average distance from a central station to these peripheral accessible stations is relatively large, which lowers the overall closeness centrality scores. This is expected, since many accessible nodes are not directly connected to one another. Overall, the outer sections of the system contain fewer accessible stations, leading to lower closeness centrality across the accessible network.

\begin{table}[!t]
\caption{The ten nodes in the NYC Metro network with the highest centrality measures and whether they are accessible.}
\vspace{-10pt}
\begin{adjustwidth}{-.1in}{.05in}
\begin{subtable}[h]{.4\textwidth}
\centering
\caption{Betweenness}
\center
\def\arraystretch{1.25}
\hspace*{-20mm}
\begin{tabular}{|p{5cm}|c|c|}
\hline
Node Name & Accessible?&  Betweenness\\
\hline\hline
Atlantic Ave - Barclay's Center &Y&0.359\\\hline
Times Square - 42nd Street &Y&0.325\\\hline
Grand Central - 42nd Street &Y& 0.248\\\hline
59th Street&Y& 0.224\\\hline
Union Square - 14th Street&Y&0.201\\\hline
Broadway Junction &N& 0.195\\\hline
Hoyt Schermerhorn &N&0.175\\\hline
59th Street - Columbus Circle&Y&0.165\\\hline
Canal Street (6) &Y& 0.164\\\hline
Nostrand Ave (A) &N&0.160\\
\hline
\end{tabular}
\end{subtable}%
\hfill
\begin{subtable}[h]{0.4\textwidth}
\centering
\caption{Closeness}
\center
\def\arraystretch{1.25}
\hspace*{-10mm}
\begin{tabular}{|p{4.8cm}|c|c|}
\hline
Node Name & Accessible?&  Closeness\\
\hline\hline
59th Street&Y&0.132\\\hline
Grand Central - 42nd Street &Y& 0.131\\\hline
Atlantic Ave - Barclay's Center &Y&0.131\\\hline
Union Square - 14th Street &Y&0.130\\\hline
36th Street (N) & N&0.129\\\hline
Times Square - 42nd Street &Y&0.128\\\hline
Canal Street (6) &Y&0.128\\\hline
Herald Square - 34th St&Y&0.127\\\hline
DeKalb Ave&Y& 0.125\\\hline
Grand Street&Y&0.123\\
\hline
\end{tabular}
\end{subtable}
\begin{flushleft} rounded to the third decimal place.
\end{flushleft}
\label{tab:nyctot}
\end{adjustwidth}
\end{table}

\subsubsection{Total Network}
Next, we compare the centrality measures of the accessible network with those of the total network. This allows us to determine whether the stations ranked in the top ten of the accessible network retain similar centrality values in the full network. We also examine how many of the top ten stations in the total network are accessible. The full New York City Metro network consists of $436$ nodes and $527$ edges, giving a cyclomatic number of $\mu = 92$. The accessible network retains $38$ of these $92$ independent cycles ($41\%$), a smaller share than in London.

\begin{figure}[b!]
\centering
\includegraphics[width=170mm]{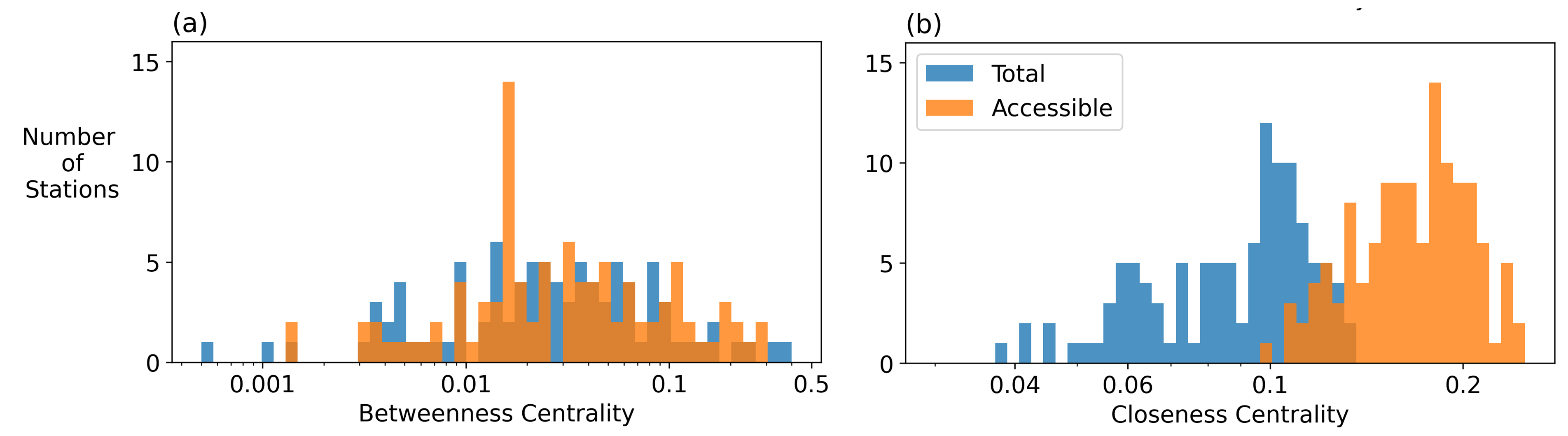}
\caption{Number of stations in the full and accessible networks by (a) betweenness and (b) closeness centrality. Raw values are not directly comparable between the two networks.}
\label{fig:nyccomall}
\end{figure}

\begin{figure}[t!]
\centering
\includegraphics[width=170mm]{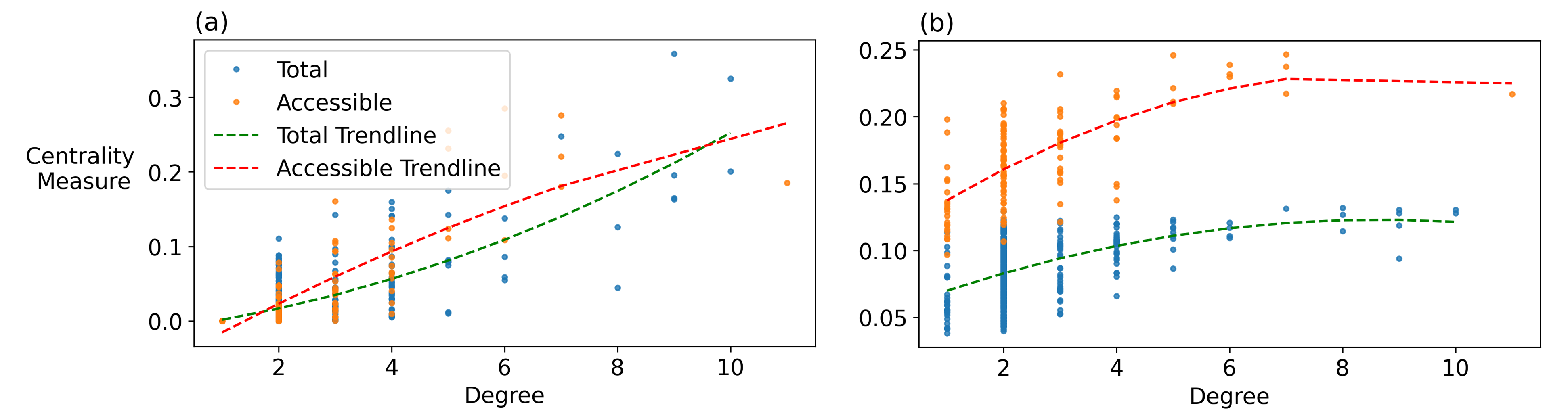}
\caption{(a) Betweenness and (b) closeness centrality values compared to degree of the nodes for the NYC Metro network.  Trendlines are added to visualize correlation between these centrality measures and degree. Raw values are not directly comparable between the two networks.}
\label{fig:nyccomgrap}
\end{figure}

As shown in Table \ref{tab:nyctot}(a), seven of our top ten stations are accessible. Atlantic Ave - Barclay’s Center has the highest betweenness centrality in the complete network, with a value of $0.359$, which is larger than its betweenness centrality in the accessible network, which is $0.185$.  This change occurs because the station is more strongly connected to non-accessible stations rather than accessible ones.  Thus, this station plays a more central role in the complete network than in the accessible sub-network. More generally, restricting the system to accessible stations reorders betweenness more in New York City than in London. The rank correlation between a station's betweenness in the accessible and full networks is $\rho = 0.69$, compared with $0.86$ in London, while for closeness it is $\rho = 0.84$ ($p < 0.001$ in all cases, $n = 125$). The overall arrangement of central and peripheral stations is therefore largely preserved, but which stations lie on the most shortest paths changes more when inaccessible stations are removed.

In the New York City full network, stations with high betweenness centrality such as Atlantic Avenue - Barclays Center, Times Square - $42$nd Street, and Grand Central - $42$nd Street serve as major transfer gateways linking multiple boroughs and lines. Their high rankings in both the total and accessible networks suggest that accessibility investments have reached several of the system’s most important nodes. However, the appearance of non-accessible stations such as Broadway Junction and Hoyt Schermerhorn among the most central full-network stations indicates that some critical transfer opportunities remain unavailable to riders requiring step-free access.

Panel (a) of Fig.~\ref{fig:nyccomall} shows that the betweenness centrality values for stations in both networks are generally low, with many clustered around approximately $0.05$. Panel (b) shows the corresponding distributions of closeness centrality, which, as in London, are not compared directly between the two networks.

\begin{figure}[!b]
\centering
\includegraphics[width=80mm]{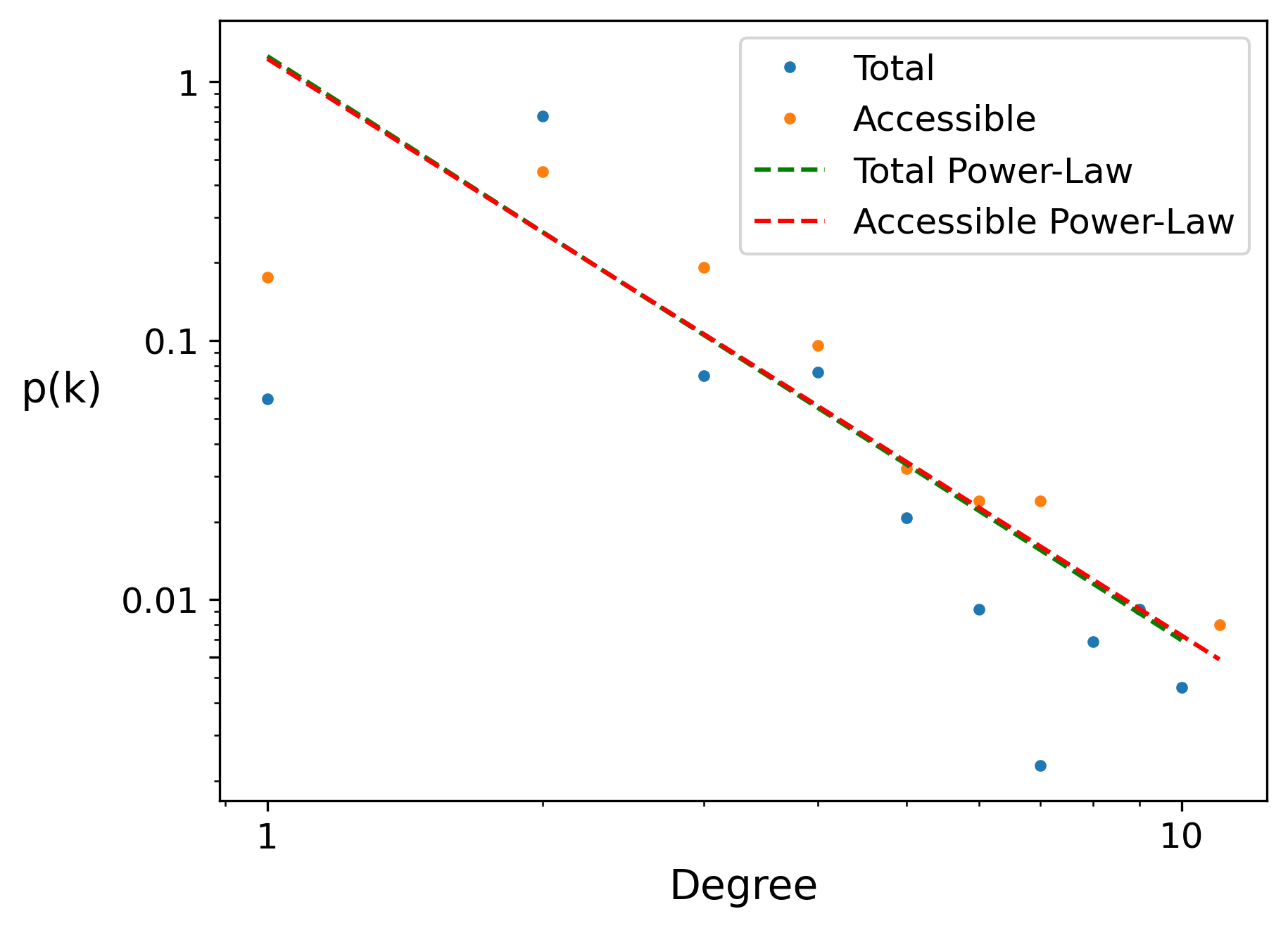}
\caption{Degree distribution of the New York City accessible and full networks. Shown on logarithmic axes with power-law fits.}
\label{fig:nycnpl}
\end{figure}

From Table \ref{tab:nyctot}, we observe that nine of the top ten stations with the highest closeness centrality in the total network are accessible. The stations closest to the rest of the full system are therefore almost all accessible, consistent with the concentration of accessible stations at major interchanges.

Next, we examine the relationship between centrality measures and node degree in both the accessible and total networks. As shown in Fig. \ref{fig:nyccomgrap}(a), betweenness centrality exhibits a positive correlation with degree in both networks. For nodes with lower degree, betweenness centrality values are similar across the accessible and total networks. However, as degree increases, the centrality values diverge somewhat, though the overall positive correlation between betweenness centrality and degree remains evident in both networks.

In contrast, the comparison of closeness centrality, shown in Fig. \ref{fig:nyccomgrap}(b), follows a similar pattern to that observed in the London Underground network (Fig. \ref{fig:loncomgrap}(b)). The trendlines indicate that, in both networks, closeness centrality values approach an upper limit as node degree increases. Overall, a positive correlation between closeness centrality and degree is evident in both networks.

Lastly, we investigate whether our networks exhibit scale-free behavior, characterized by a power-law distribution.  As shown in Fig.~\ref{fig:nycnpl}, the degree distribution of both networks roughly follow a power law, with exponents $\gamma_{A2}=2.227$ for the accessible network and $\gamma_{T2}=2.255$ for the complete network. As in London, the narrow range of degrees means these exponents indicate power-law-like behavior rather than establishing a power law \cite{Clauset09}. These results are consistent with previous studies that indicate that public transit networks modeled in $\mathbb{L}$-space display scale-free properties \cite{Derrible11, Sen03, Sien05, Ferber07, Ferber09}.

\begin{figure}[!b]
\centering
\includegraphics[width=100mm]{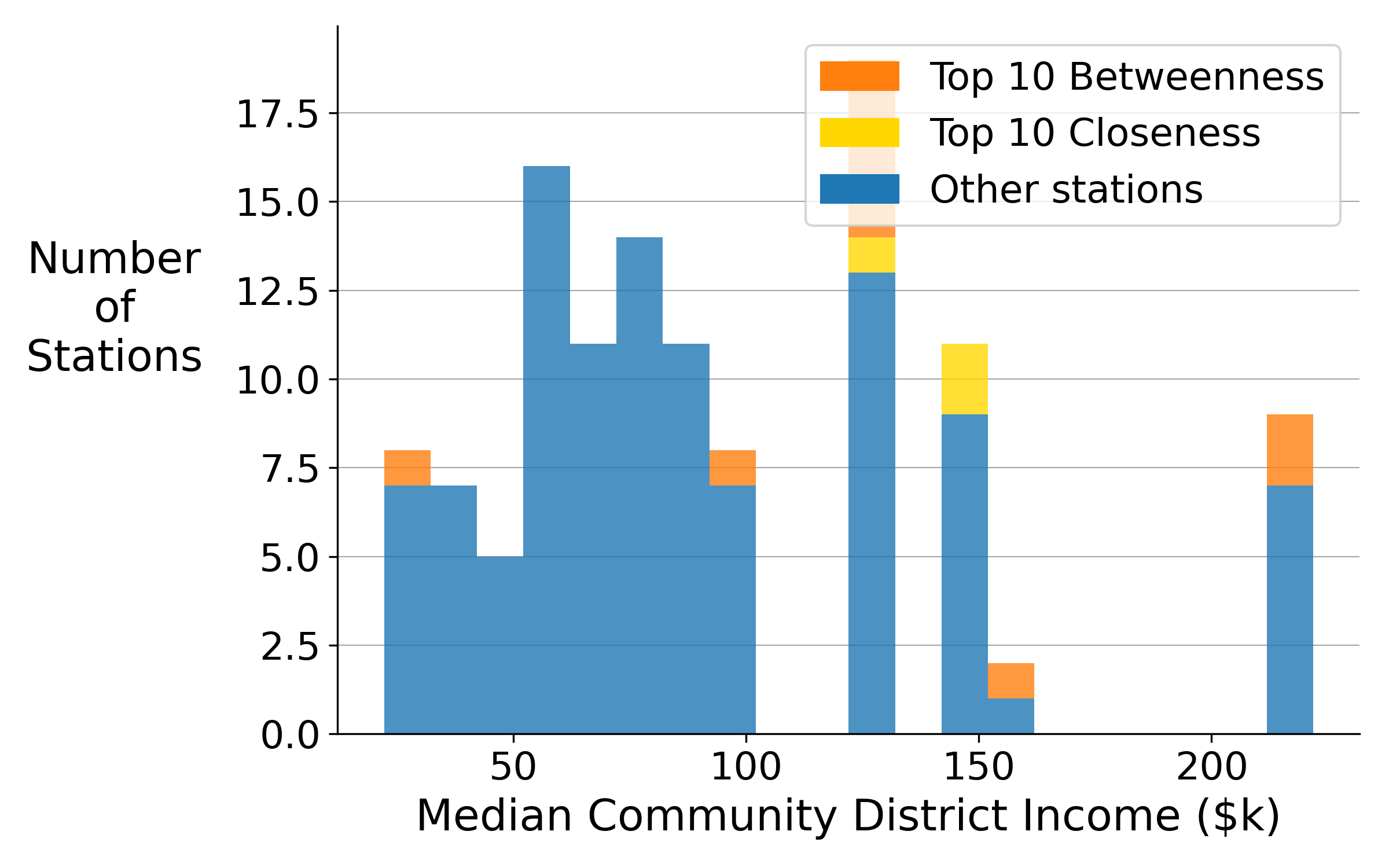}
\caption{Median annual income, per thousand dollars ($\$k$), per community district and the number of accessible stations in each community district. The stations in our top ten list of betweenness and closeness centralities and their median annual income are indicated in orange and yellow, respectively.}
\label{fig:nycinc}
\end{figure}

\subsubsection{Income and Tourism Results}
As shown in Fig. \ref{fig:nycinc}, accessible stations are distributed across community districts with varying income levels, and the relationship between income and accessibility does not appear uniform across the city. Some higher-income districts contain multiple accessible stations, but accessible stations are also present in several moderate- and lower-income areas. This suggests that income alone is unlikely to explain the spatial pattern of accessibility.

\begin{figure}[t!]
\centering
\includegraphics[width=170mm]{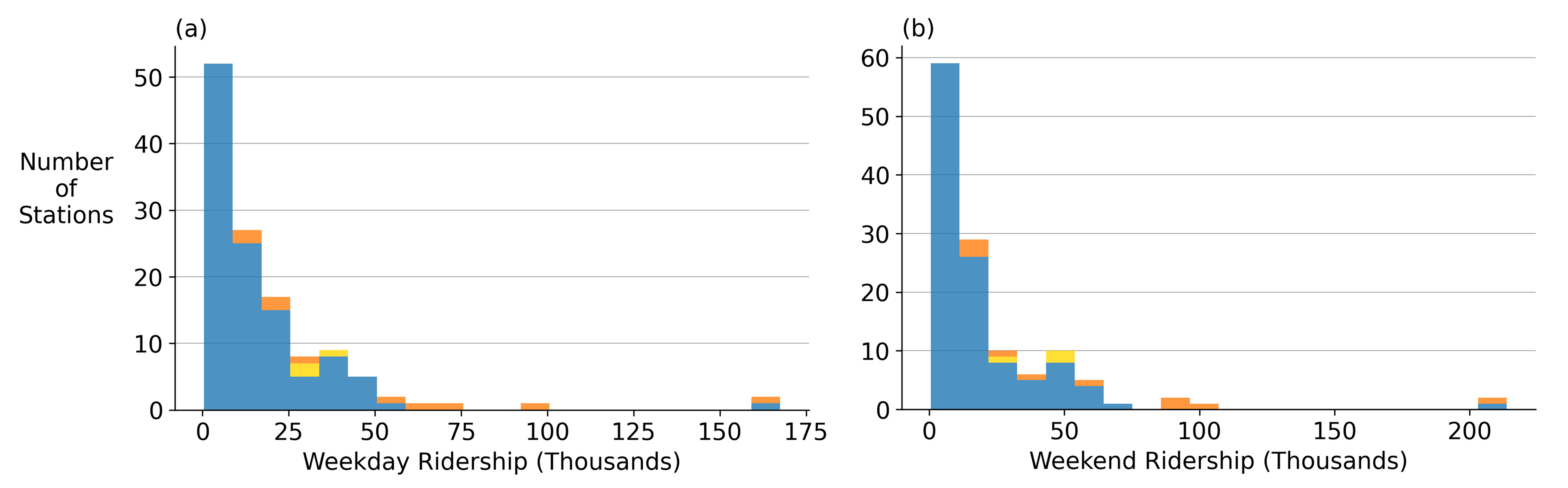}
\caption{Number of accessible stations by (a) weekday and (b) weekend ridership. The accessible stations in our top ten lists of betweenness and closeness centralities and their ridership are indicated in orange and yellow, respectively.}
\label{fig:nycpop}
\end{figure}

Figure \ref{fig:nycpop} shows the weekday and weekend ridership of each accessible station. Most accessible stations have modest ridership, but several of the stations with the highest betweenness and closeness centrality are among the busiest accessible stations, particularly near major employment centers and major interchanges. Because the figure includes only accessible stations, it does not show whether accessible stations are busier than inaccessible ones, and ridership alone cannot establish why these stations were made accessible. Weekend ridership may also reflect destinations with commercial, recreational, or visitor demand, although these measures do not isolate tourism specifically.

Overall, the New York City results suggest that accessible stations are often concentrated in high-activity parts of the network, especially around Manhattan and major interchange locations. However, accessibility patterns may reflect multiple overlapping influences, including historical station design, capital investment priorities, engineering constraints, and ridership demand \cite{MTAada23}.

\subsection*{Comparison with random allocation}
To test whether the placement of accessible stations differs from what would be expected by chance, we compared the observed accessible stations in each city with $1000$ random sets of the same size (Table \ref{tab-null}). In London, the accessible stations have a higher mean betweenness centrality than $98.9\%$ of uniform random sets ($p=0.024$). Under the degree-matched null model, in which the random sets contain the same number of interchange stations as the observed set, the observed value falls at the $97$th percentile and the difference is no longer significant at the 5\% level ($p=0.062$). The high betweenness of London's accessible stations is therefore partly explained by the selection of interchange stations.

\begin{table}[!t]
\begin{adjustwidth}{-.8in}{0in}
\centering
\caption{Accessible stations compared with 1000 random allocations of the same size.
Mean full-network betweenness centrality of the accessible stations and the share of all stations within two stops of an accessible station ($C_2$), compared with random sets of the same size drawn under the uniform and degree-matched null models. Null values are the mean and the 95\% interval (2.5th to 97.5th percentiles) of the null distribution. The percentile is the share of random sets with a lower value than the observed set, and $p$ is the two-sided empirical $p$-value. Results for all statistics are given in Appendix~\ref{app:rat}.}
\label{tab-null}
\def\arraystretch{1.25}
\begin{tabular}{|l|c|c|c|c|c|}
\hline
 & & \multicolumn{2}{c|}{Uniform null} & \multicolumn{2}{c|}{Degree-matched null}\\
\cline{3-6}
Statistic & Observed & Mean [95\% interval] & Percentile ($p$) & Mean [95\% interval] & Percentile ($p$)\\
\hline\hline
\multicolumn{6}{|l|}{\textbf{London}}\\\hline
Mean betweenness & 0.0428 & 0.0355 [0.0288, 0.0420] & 98.9 (0.0240) & 0.0377 [0.0319, 0.0428] & 97.0 (0.0619)\\\hline
$C_2$ & 0.9228 & 0.9697 [0.9436, 0.9911] & 0.3 (0.0120) & 0.9722 [0.9465, 0.9911] & 0.0 (0.0020)\\\hline\hline
\multicolumn{6}{|l|}{\textbf{New York City}}\\\hline
Mean betweenness & 0.0395 & 0.0264 [0.0208, 0.0331] & 100.0 (0.0020) & 0.0376 [0.0326, 0.0425] & 75.8 (0.4855)\\\hline
$C_2$ & 0.9106 & 0.8742 [0.8348, 0.9106] & 97.0 (0.0619) & 0.8744 [0.8349, 0.9106] & 96.4 (0.0739)\\\hline
\end{tabular}
\end{adjustwidth}
\end{table}

Coverage shows the opposite pattern. Only $92.3\%$ of London stations lie within two stops of an accessible station, compared with an average of $97.0\%$ under uniform random allocation ($p=0.012$), and the difference becomes stronger under the degree-matched null model ($p=0.002$). The same number of accessible stations placed at random would leave fewer gaps in the network. This is consistent with the clustering of London's accessible stations along lines that are step-free throughout, such as the Elizabeth line and the DLR, where an accessible station is often next to other accessible stations and so adds little to coverage. The mean distance from an inaccessible station to the nearest accessible station gives the same result as $C_2$ (Appendix~\ref{app:rat}).

In New York City, the accessible stations have a higher mean betweenness centrality than every uniform random set ($p=0.002$). They also have a much higher mean degree than random sets ($2.98$ compared with $2.42$, Appendix~\ref{app:rat}), so accessible stations are disproportionately interchange stations. Under the degree-matched null model, the betweenness advantage disappears ($75.8$th percentile, $p=0.486$). The high betweenness of New York City's accessible stations is therefore fully explained by the selection of interchange stations.

Coverage in New York City is similar to or slightly better than random. In total, $91.1\%$ of stations lie within two stops of an accessible station, compared with an average of $87.4\%$ under uniform random allocation, but the difference is not significant at the $5\%$ level ($p=0.062$ under the uniform null model and $p=0.074$ under the degree-matched null model). The mean distance from an inaccessible station to the nearest accessible station is lower than under uniform random allocation ($p=0.042$) but not under the degree-matched null model (Appendix~\ref{app:rat}).

The two cities therefore show two distinct patterns of allocation. In New York City, accessible stations are concentrated at interchanges, which fully accounts for their central position, and they cover the network about as well as a random placement of the same number of stations. In London, accessible stations are more central than random allocations, partly because many of them are interchange stations, and they remain somewhat more central than degree-matched random sets, although this difference is not significant at the $5\%$ level. They are also clustered along lines that are step-free throughout, leaving more stations far from an accessible station than a random placement would. These patterns cannot by themselves be attributed to differences in policy between the two systems, but they show that the same random-allocation test distinguishes different allocation strategies and can be applied to any transit network with station-level accessibility data.

\section{Discussion}\label{sec:conc}
In this study, we analyzed the accessible-station networks of two major public transit systems: the London Underground and the New York City Metro.  A key comparative finding is that the two systems exhibit different patterns in how accessibility is distributed across their networks. London has a substantially larger proportion of accessible stations, with $162$ accessible nodes represented in the accessibility network compared with $125$ in New York City, and accessible stations make up approximately $48\%$ of the nodes in the full London network, compared with $29\%$ in New York City \cite{TFLel,NYCaccsize}.

In contrast, accessible stations in New York City are fewer in number and are more concentrated around major transfer points and central Manhattan corridors. These structural differences are reflected in the network metrics. London's accessible network retains a larger share of the independent cycles of its full network ($56\%$ versus $41\%$), meaning that more of the system's alternative routes remain available to riders who rely on accessible stations. The two centrality measures also behave differently in the two cities. In New York City, betweenness and closeness rankings within the accessible network are moderately correlated ($\rho = 0.58$), so the stations that lie on the most step-free shortest paths are largely those closest to the rest of the accessible network, consistent with a compact accessible core. In London the correlation is weak ($\rho = 0.28$), reflecting long step-free corridors in which important transfer points are not necessarily central in distance.

London's accessible network also has a higher peak betweenness centrality. Roughly half of pairs of other accessible stations are connected by shortest paths through Stratford ($0.488$), compared with $0.285$ for Grand Central--$42$nd Street in New York City, so step-free routes in London depend more heavily on their most central interchange. These differences may partly reflect the different ways the two systems have acquired accessible stations. A substantial share of London's accessible network lies on lines built step-free from the outset, such as the Elizabeth line and the DLR \cite{TFLel}, whereas accessibility in New York City has come largely through retrofitting existing stations under a legal mandate, where site conditions constrain which stations can be upgraded \cite{MTAada23}. Despite these differences, accessible stations in both systems include many of the network's major interchanges. Comparison with random allocations shows that in London this partly reflects the selection of interchange stations, and that the same allocation leaves coverage gaps that a random placement of the same number of stations would not. In New York City, the central position of accessible stations is fully explained by their concentration at interchanges, and the same stations cover the network about as well as a random placement would.

The centrality analysis also shows how each accessible network is organized. Stations with high betweenness centrality lie on a large share of the shortest step-free paths between accessible stations, so they act as the main transfer points of the accessible network. In both cities, several of the most central accessible stations are major interchange hubs. Comparing each accessible station's centrality in the accessible and full networks shows how restricting the system to accessible stations changes these roles. In London, betweenness rankings are largely preserved ($\rho = 0.86$), so the stations that structure shortest paths in the full system also structure step-free shortest paths. In New York City they are reordered more ($\rho = 0.69$), so step-free shortest paths pass through a somewhat different set of stations than shortest paths in the full system, even though the overall arrangement of central and peripheral stations is preserved ($\rho = 0.84$ for closeness).

Beyond the findings for London and New York, the random-allocation framework provides a methodological approach that can be applied to other transit systems for which station-level accessibility data are available. By comparing the observed distribution of accessible stations with both random and degree-matched allocations, this approach helps distinguish whether the structural positions of accessible stations reflect their tendency to be located at highly connected stations or patterns that persist beyond degree alone. Thus, the framework offers a way to evaluate the structural placement of accessible stations across transit networks, rather than limiting the analysis to comparisons of the number or proportion of accessible stations.

We also examined the socioeconomic context of accessible stations, and these comparisons are descriptive rather than causal. In both cities, accessible stations frequently occur in areas with substantial daytime activity, employment concentration, or major destination functions. By contrast, the relationship between accessibility and local income measures is mixed and should be interpreted cautiously given the use of aggregated geographic units.

Several avenues exist to make the network model more realistic. Incorporating weighted edges could account for differences in service frequency, line overlap, or passenger volumes. Modeling the networks as directed graphs could capture stations that are accessible in only one direction or require more complex transfer paths. Future work could also incorporate operational reliability, such as elevator outages, maintenance disruptions, or real-time service changes, which may substantially affect practical accessibility beyond nominal station status \cite{NYCaccsize}.

This study also has limitations. The socioeconomic variables were measured using borough- and district-level administrative units, which may not fully represent the catchment area of individual stations. Some supporting datasets were not available for identical years across variables or cities. In addition, classifying stations as accessible does not guarantee full real-world usability, particularly where elevators are temporarily unavailable or platform gaps remain. Because only two systems are analyzed, the results cannot separate the effect of a legal mandate, such as the ADA key-station requirement in New York City, from other influences on station selection. Applying the random-allocation test to additional systems, including systems with and without such mandates, is a natural direction for future work.

Overall, the findings highlight that accessibility is not solely a question of how many stations are accessible, but also where those stations are located within the network. Evaluating accessibility therefore requires considering both the structural importance of accessible stations and how effectively they provide coverage across the rest of the network. A station may occupy a highly central position and serve as an important transfer point, but this does not necessarily mean that accessible stations are distributed in a way that minimizes gaps in network coverage. This distinction is evident in the comparison of London and New York City where, although accessible stations in both systems are concentrated at major interchanges, London’s accessible stations leave more of the network beyond two stops of an accessible station than would be expected under random allocation. Even when many stations are accessible, the absence of step-free access at structurally important interchanges, such as Baker Street in London and Broadway Junction in New York City, limits which connections in the system remain available to riders who rely on step-free infrastructure. Our results suggest that improving accessibility involves not only additional station upgrades, but also attention to network-wide connectivity, reliability, and spatial coverage.

\section*{Data Availability Statement} 
Data sets for the London Underground are available in the Zenodo Repository: \url{https://doi.org/10.5281/zenodo.19824433}. Data sets for the New York Metro are available in the Zenodo Repository: \url{https://doi.org/10.5281/zenodo.19824493}.

Data regarding workday population, tourist population, and income for London are available at London DataStore: \url{https://data.london.gov.uk/}. Median household income data for New York City were obtained from the U.S. Census Bureau’s American Community Survey 5-year estimates (2021, table B19013) through the Census API. Station ridership data for New York City were compiled by the Metropolitan Transportation Authority and are available at \url{https://www.mta.info/agency/new-york-city-transit/subway-bus-ridership-2024}.
\section*{Acknowledgments} 
This research received no specific grant from any funding agency, commercial or not-for-profit sectors. The author would like to thank Naoki Masuda for his guidance and assistance with editing.
\bibliographystyle{unsrt}
\bibliography{Betz_Access_Refs}
\begin{appendices}
\section{Random-allocation Test Results}\label{app:rat}

\begin{table}[!ht]
\centering
\caption{Full results of the random-allocation test for London and New York City.
Observed values for the accessible stations compared with $1,000$ random sets of the same size drawn under the uniform and degree-matched null models. All statistics are computed on the full network. The $95\%$ interval spans the $2.5$th to $97.5$th percentiles of the null distribution, the percentile is the share of random sets with a lower value than the observed set, and $p$ is the two-sided empirical $p$-value. $H$ is the mean distance in stops from an inaccessible station to the nearest accessible station, and $C_2$ is the share of stations within two stops of an accessible station.}
\def\arraystretch{1.2}
\begin{tabular}{|l|c|c|c|c|c|}
\hline
Statistic & Observed & Null mean & 95\% null interval & Percentile & $p$\\
\hline\hline
\multicolumn{6}{|l|}{\textbf{London}, uniform null}\\\hline
Mean betweenness & 0.0428 & 0.0355 & [0.0288, 0.0420] & 98.9 & 0.0240\\\hline
Mean closeness & 0.0798 & 0.0827 & [0.0805, 0.0850] & 0.3 & 0.0080\\\hline
Mean degree & 2.4410 & 2.3574 & [2.2236, 2.4845] & 88.9 & 0.2238\\\hline
$H$ (stops) & 1.5966 & 1.3438 & [1.2216, 1.5399] & 99.0 & 0.0220\\\hline
$C_2$ & 0.9228 & 0.9697 & [0.9436, 0.9911] & 0.3 & 0.0120\\\hline
\multicolumn{6}{|l|}{\textbf{London}, degree-matched null}\\\hline
Mean betweenness & 0.0428 & 0.0377 & [0.0319, 0.0428] & 97.0 & 0.0619\\\hline
Mean closeness & 0.0798 & 0.0825 & [0.0804, 0.0843] & 0.4 & 0.0100\\\hline
Mean degree & 2.4410 & 2.3949 & [2.3478, 2.4410] & 97.1 & 0.0599\\\hline
$H$ (stops) & 1.5966 & 1.3204 & [1.2102, 1.4830] & 99.8 & 0.0060\\\hline
$C_2$ & 0.9228 & 0.9722 & [0.9465, 0.9911] & 0.0 & 0.0020\\\hline\hline
\multicolumn{6}{|l|}{\textbf{New York City}, uniform null}\\\hline
Mean betweenness & 0.0395 & 0.0264 & [0.0208, 0.0331] & 100.0 & 0.0020\\\hline
Mean closeness & 0.0906 & 0.0863 & [0.0832, 0.0894] & 99.7 & 0.0080\\\hline
Mean degree & 2.9760 & 2.4154 & [2.2478, 2.6320] & 100.0 & 0.0020\\\hline
$H$ (stops) & 1.6045 & 1.7874 & [1.6109, 2.0482] & 1.9 & 0.0420\\\hline
$C_2$ & 0.9106 & 0.8742 & [0.8348, 0.9106] & 97.0 & 0.0619\\\hline
\multicolumn{6}{|l|}{\textbf{New York City}, degree-matched null}\\\hline
Mean betweenness & 0.0395 & 0.0376 & [0.0326, 0.0425] & 75.8 & 0.4855\\\hline
Mean closeness & 0.0906 & 0.0891 & [0.0865, 0.0915] & 88.1 & 0.2398\\\hline
Mean degree & 2.9760 & 2.8900 & [2.7840, 2.9840] & 93.7 & 0.1279\\\hline
$H$ (stops) & 1.6045 & 1.7794 & [1.5787, 2.0900] & 6.0 & 0.1279\\\hline
$C_2$ & 0.9106 & 0.8744 & [0.8349, 0.9106] & 96.4 & 0.0739\\
\hline
\end{tabular}
\end{table}

\end{appendices}
\end{document}